\documentclass[%
 preprint,
superscriptaddress,
 amsmath,amssymb,
 aps,
aip,
floatfix,
]{revtex4-1}

\usepackage{graphicx}
\usepackage{dcolumn}
\usepackage{bm}

\usepackage{color,soul}
\usepackage{epstopdf}
\AppendGraphicsExtensions{.tif}

\begin{document}


\title{Interfacial Structure of SrZr$_{x}$Ti$_{1-x}$O$_3$ films on Ge }
\author{Tongjie Chen}
 \affiliation{Department of Physics, North Carolina State University, Raleigh, NC, 27695 USA}

\author{Kamyar Ahmadi-Majlan}
 \affiliation{Department of Physics,University of Texas at Arlington, Arlington, TX, 76019 USA}

\author{Zheng Hui Lim}
 \affiliation{Department of Physics,University of Texas at Arlington, Arlington, TX, 76019 USA}

\author{Zhan Zhang}
 \affiliation{Advanced Photon Source, Argonne, IL, 76019 USA}

\author{Joseph H. Ngai}
 \affiliation{Department of Physics,University of Texas at Arlington, Arlington, TX, 76019 USA}

\author{Alexander F. Kemper}
 \affiliation{Department of Physics, North Carolina State University, Raleigh, NC, 27695 USA}
 
\author{Divine P. Kumah}
\email{dpkumah@ncsu.edu}
 \affiliation{Department of Physics, North Carolina State University, Raleigh, NC, 27695 USA}

\date{\today}

\begin{abstract}
The interfacial structure of SrZr$_{x}$Ti$_{1-x}$O$_3$ films grown on semiconducting Ge substrates are investigated by synchrotron X-ray diffraction and first-principles density functional theory. By systematically tuning Zr content x, the effects of bonding at the interface and epitaxial strain on the physical structure of the film can be distinguished. The interfacial perovskite layers are found to be polarized as a result of cation-anion ionic displacements perpendicular to the perovskite/semiconductor interface. We find  a correlation between the observed buckling and valence band offsets at the SrZr$_{x}$Ti$_{1-x}$O$_3$/Ge interface. The theoretical valence band offsets for the polar structures are in agreement with reported X-ray photoelectron spectroscopy measurements. These results have important implications for the integration of functional oxide materials with established semiconductor based technologies.

\end{abstract}

\maketitle

The monolithic integration of complex transition metal-oxides on semiconductors provides a route to introduce additional material functionalities to semiconductor-based technologies. \cite{zubko2011interface,mckee2001physical,eisenbeiser2000field,xiong2014active,lorenz20162016,mazet2015review,zhang2003atomic,reiner2010crystalline}  Complex oxides exhibit short electronic length-scales over which material behavior can be altered, and thus have great potential for use in highly-scaled device technologies. In the ultra-thin limit of these complex oxide materials (1-10nm), the coupling of electronic and structural order parameters leads to emergence of novel physical properties not found in the bulk constituent materials.\cite{ngai2014correlated,kumah2016engineered,kolpak2010interface,ngai2017electrically,ahmadi2018tuning} In this regard, (100)-oriented surfaces of diamond-cubic semiconductors impose mechanical and electrostatic boundary conditions that can induce novel changes in the physical and electronic structure of ultra-thin epitaxial oxides. 

The interface between SrTiO$_3$ (STO) and Si serves as an archetype system to understand the effects of mechanical and electrostatic boundary conditions imposed by the substrate. Studies of the STO/Si system reveal the presence of a fixed polarization that is initiated by charge transfer at the interface and enhanced by epitaxial strain.\cite{kolpak2012interface,kumah2010atomic,reiner2010crystalline} . However, separating the effects of interfacial bonding from epitaxial strain on physical structure is challenging through a study of the STO/Si system alone. Furthermore, elucidating the effects of the polarization on band alignment, which is a key property governing material functionality, has yet to be experimentally explored.    

In comparison, the epitaxial SrZr$_{x}$Ti$_{1-x}$O$_3$ (SZTO)/Ge system provides a model system for detangling the role of strain and electrostatic boundary conditions on the interfacial structural and electronic properties of perovskite oxide/semiconductor interfaces. By varying the Zr content \textit{x}, from 0 to 1, the strain in the oxide layer can be continuously varied from tensile to compressive strain, while maintaining the chemical composition of the STO/Si interface. Additionally, the oxide band gap can be tuned from 3.2 eV for STO to 5.6 eV for SrZrO$_3$ (SZO),which enables the band alignment to be altered from type II (staggered type) to type I (straddling type)\cite{jahangir2015band,chambers2017effects,lim2017structural}. X-ray photoelectron spectroscopy (XPS) and transport measurements have demonstrated a change in the band alignment from type II (staggered type) to type I (straddling type) as the Zr content is increased.\cite{jahangir2015band}Thus, the effects of interfacial structure on band-alignment can also be determined. 


In this study, we use a combination of first-principles density functional theory (DFT) and high resolution synchrotron X-ray diffraction measurements of crystal truncation rods to determine the atomic-scale structure of nominally 2.5 unit cell (uc) thick SZTO  films grown on Ge as a function of the Zr content. Symmetry-breaking polar displacements of oxygen anions relative to the metal cations normal to the plane of the film are found from the analysis of crystal truncation rods in the SZTO films for all Zr content. An upward polarization is observed at the interface and throughout the film for films under compressive strain. In contrast, a reversal in the direction of the polarization near the top surface is observed for films under large tensile strain.  First-principles density functional calculations are used to confirm the measured interfacial structures and elucidate the effects of interfacial polarizations found in the films on the band alignments at the SZTO/Ge interface.  We find that the calculated band-offsets from DFT are in agreement with previously reported photoelectron spectroscopy measurements. \cite{chambers2017effects,lim2017structural} 

SZTO films were grown on Ge(001) substrates using molecular beam epitaxy.\cite{jahangir2015band}  A series of films 2.5 uc thick were grown with 0\% Zr(SZTO x=0), 30\% Zr (SZTO x=0.3) and 100\% Zr(SZTO x=1.0). 
Transmission electron microscope studies of films grown under these conditions show atomically sharp interfaces without the formation of interfacial GeO$_x$.\cite{jahangir2015band,lim2017structural} The first oxide layer adjacent to the interface is a SrO$_y$ layer as expected from the growth sequence. The nominal layer stacking along the growth direction is  Ge/SrO$_y$ /(Zr,Ti)O$_2$/SrO/(Zr,Ti)O$_2$/SrO. The [001] axis of the oxide film is parallel to the [001] axis of the Ge substrate and the [100] perovskite axis is parallel to the Ge [110] axis (i.e. the perovskite unit cell is rotated 45$^o$ in-plane with respect to the Ge [001] axis). 

\begin{figure*}[ht]
\includegraphics[scale=0.55]{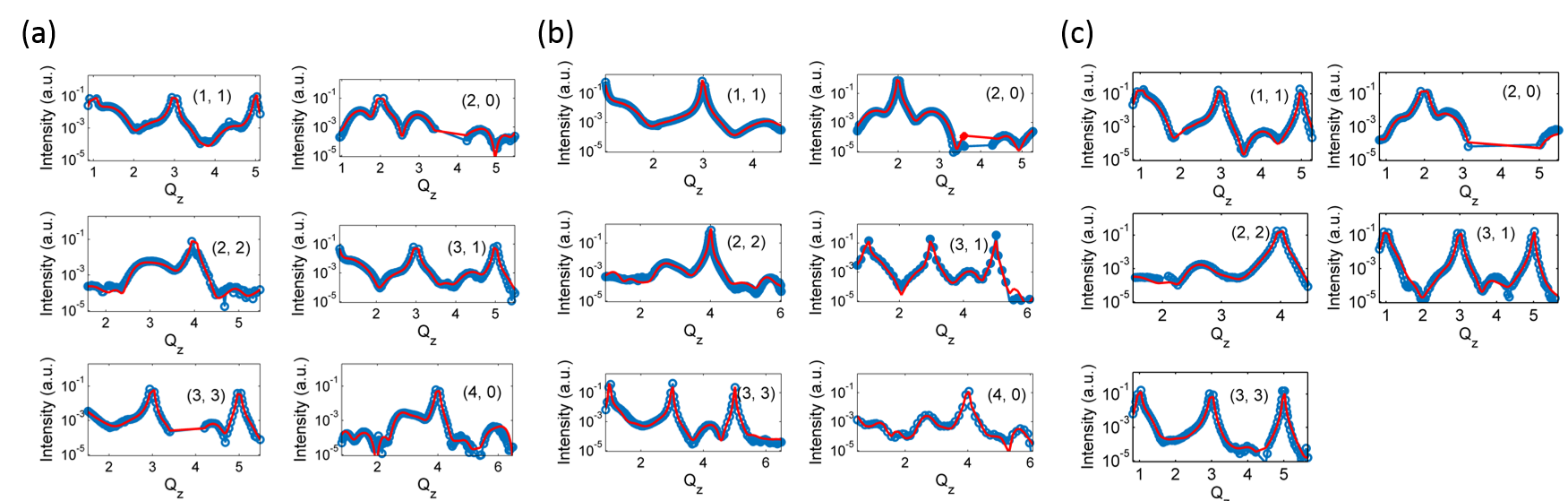} 
\caption{Comparison of measured crystal truncation rods(blue circles) and associated fits (red lines) for 2.5 unit cell thick (a)  SrTiO$_3$ on Ge (b) SrTi$_{0.3}$Zr$_{0.7}$O$_3$ on Ge and (c) SrZrO$_3$ on Ge. The reciprocal space values are in units of the bulk Ge reciprocal lattice units.  }
\label{fig:STO_fit}
\end{figure*}

To determine the positions of the cations (Sr,Zr,Ti) and the oxygen ions with sub-Angstrom resolution, the  films were capped \textit{in-situ} after growth with either amorphous Si or amorphous Ti layers for \textit{ex-situ} synchrotron diffraction experiments at the Advanced Photon Source.\cite{kumah2010atomic} Crystal truncation rod (CTR) measurements were performed at the 33-ID beamline with an incident photon energy of 16 keV ($\lambda$=0.774 \AA{}). CTR measurements were performed along directions in reciprocal space defined by the bulk Ge reciprocal lattice units (Ge r.l.u.) for the three 2.5 unit cell samples with x=0 , x=0.3 and x=1. CTRs were measured along the (00L), (11L), (20L), (22L), (31L), (33L) and (40L) rods for $0.5<L<5.5$ Ge. r.l.u. No half-order reflections which would arise from a doubling of the perovskite unit cell and/or a dimerization of the interfacial Ge were observed.\cite{kumah2016engineered}

The atomic-scale structures of the films were determined by fitting the CTRs using the GENX genetic fitting algorithm.\cite{bjorck2007genx,kumah2010atomic,kumah2016engineered}  Figure \ref{fig:STO_fit} (a-c) shows the measured CTRs and simulated intensities of the best-fit structures for the SZTO/Ge samples for x=0, 0.3 and 1 respectively. 
 Convergence of the fits were evaluated using the crystallographic R-factor where R-factor$=\frac{\sum |F_{measured} - F_{caclulated}|}{\sum |F_{measured}|}$. $F_{measured}$ and  $F_{calculated}$ are the measured and calculated crystallographic structure factors respectively. The atomic coordinates along the z-direction, Debye-Waller factors and layer occupations were varied until convergence of the R-factor was achieved for each sample. To determine the oxygen content of the interfacial SrO$_y$ layer, we allowed the oxygen occupation parameter for this layer to vary; however, it converged on values close to 0. Hence, we conclude that the first perovskite layer is a Sr layer. This is consistent with a 1x1 structure due to the destabilization of the Ge dimerization resulting from charge transfer from the oxygen vacancy to the interfacial Ge layer.\cite{kumah2016engineered} The converged R-factors for the x=0, x=0.3 and x=1 films are 7\%, 5\% and 6\% respectively.

Theoretical structures were calculated for x=0, x=0.5 and x=1.0 using DFT for comparison with the measured structures. The DFT calculations were performed using the Quantum Espresso\cite{giannozzi2009quantum} software package with Perdew-Burke-Ernzerhof exchange functionals\cite{perdew1996generalized}. For the 50\% Ti/Zr mixture the virtual crystal approximation was utilized to construct the pseudo-potential. The wave function and density energy cutoffs were chosen to be 60 and 240 Ry, respectively, and a $13 \times 13 \times 1$ momentum grid was used. The calculations considered an Hydrogen-terminated Ge slab with 2.5 monolayers of STO where the O in the bottom-most Sr-O plane was absent based on the experimental results. We used the optimal (within DFT) lattice constant along the (001) direction of bulk Ge for the in-plane lattice constant of the calculations.  During the structural optimization, the top layer of Ge was allowed to relax, while the rest of the Ge slab was kept fixed. The DFT converged structures are shown in Figure 2 a-c. 


\begin{figure}[ht]
\includegraphics[width=0.45\textwidth]{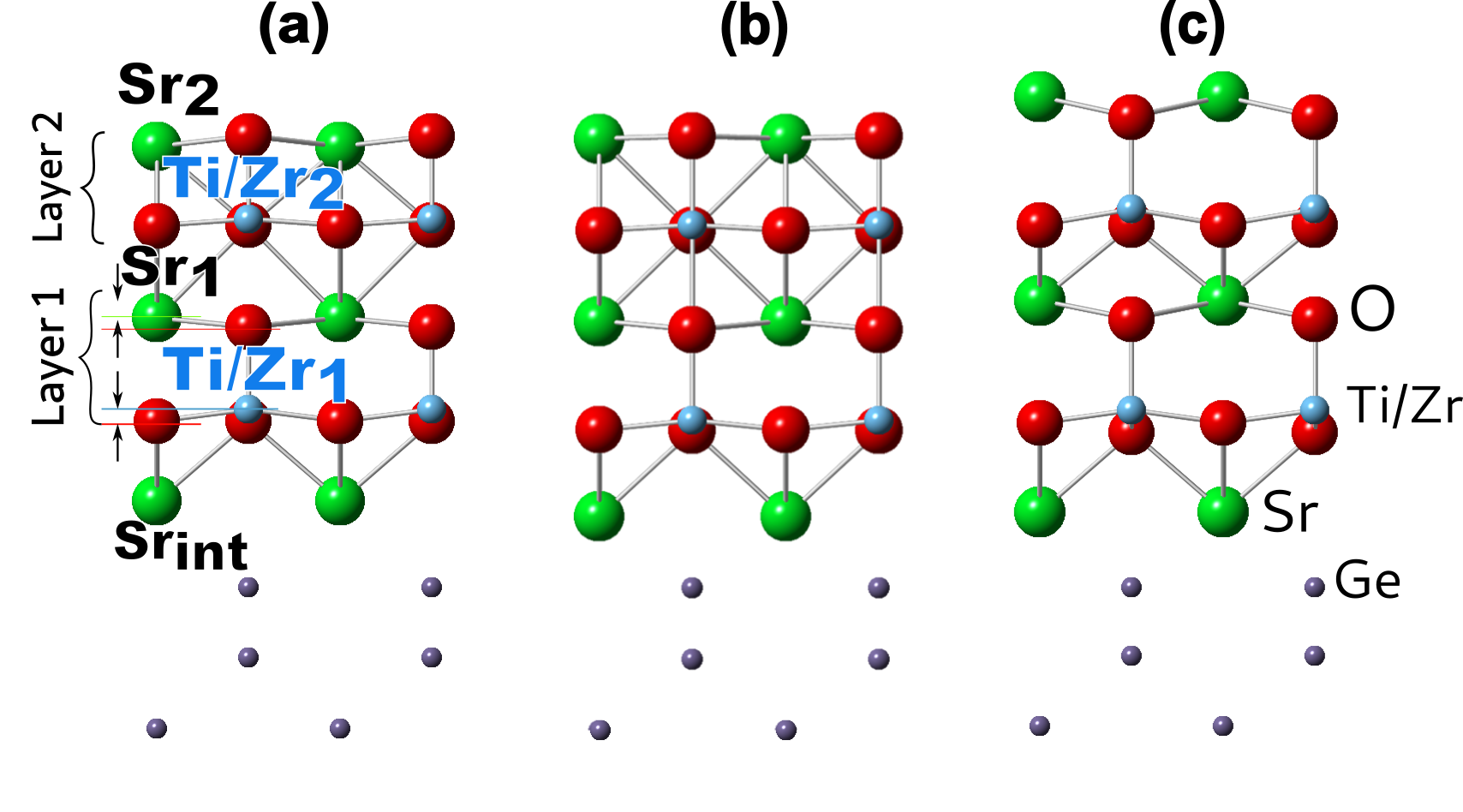}
\caption{Theoretical structures for 2.5 uc SZTO films on Ge as a function of Zr content in $SrZr_xTi_{1-x}O_3$ on Ge, for (a)$x=0$, (b)$x=0.5$ and (c)$x=1$.}
\label{fig:bond-model}
\end{figure}

The experimental layer-resolved out-of-plane lattice constants for the films were determined from the vertical spacings between the Sr planes for the best-fit structures from the CTR measurements and are summarized in Table \ref{table:1} as a function of Zr content. STO has a bulk lattice constant of 3.905 \AA{} and is thus under tensile strain when grown on the (001) Ge surface with an in-plane lattice constant of 3.99 \AA{}. A contraction of the out-of-plane lattice constant is observed as expected for the STO film. As the Zr content increases, an expansion of the lattice along the c-axis is observed as the larger Zr$^{4+}$ ions replace Ti$^{4+}$. The bulk pseudo-cubic lattice constant of (SZTO x=1.0) is 4.15 \AA{} \cite{kennedy1999high,weston2015structural}, hence, there is a transition from tensile strained (SZTO x=0) to compressively strained (SZTO x=1.0) on Ge as the Zr content is increased.

\begin{table}[h]
\begin{center}
\begin{tabular}{ | m{1.5em} | m{2.5em}| m{2.5em} |m{3em} | m{3.5em}| m{2.5em} | m{3.5em}| m{2.5em} | } 
\hline
Zr \ &	Sr$^{int}_z$-Sr$^1_z$	&	Sr$^1_z$-Sr$^2$	&	Ge$_z$-Sr$^{int} _z$ &	O$_z$-(Ti/Zr)$^1_z$	&	$O_z-Sr^2_z$	&	$O_z$-(Ti/Zr)$^2_z$	&	$O_z-Sr^2_z$ \\
\hline
\hline

0	 & 3.76	& 3.68	 & 1.50	 & -0.09	 & -0.08	 & -0.10	 & 0.18 \\
							
0.3	 & 3.93	 & 3.86	 & 1.54	 & -0.15	 & -0.35	 & -0.22	 & -0.17 \\
1.0	 & 4.28	 & 4.02	 & 1.61	 & -0.33	 & -0.43	 & -0.42	 & -0.08 \\
\hline

\end{tabular}
\caption{Experimental structure of 2.5 unit cell SrZr$_{x}$Ti$_{1-x}$O$_3$ films on Ge. The distances are in units of Angstroms. The superscripts refer to the layers in Figure 2.}
\label{table:1}
\end{center}
\end{table}

\begin{table}[h]
\begin{center}
\begin{tabular}{ | m{1.5em} | m{2.5em}| m{2.5em} |m{3em} | m{3.5em}| m{2.5em} | m{3.5em}| m{2.5em} | } 
\hline
Zr \ &	Sr$^{int}_z$-Sr$^1_z$	&	Sr$^1_z$-Sr$^2$	&	Ge$_z$-Sr$^{int} _z$ &	O$_z$-(Ti/Zr)$^1_z$	&	$O_z-Sr^2_z$	&	$O_z$-(Ti/Zr)$^2_z$	&	$O_z-Sr^2_z$ \\
\hline
\hline
0	& 3.90	& 3.61	& 1.82	& -0.25	& -0.2	& -0.16	& 0.23 \\
0.5	& 4.18	& 3.89	& 1.52	& -0.21	& -0.19	& -0.11	& 0.05 \\
1.0	& 4.59	& 4.42	& 1.62	& -0.38	& -0.40	& -0.34	& -0.43 \\

\hline

\end{tabular}
\caption{First-principles structure of 2 unit cell SrZr$_{x}$Ti$_{1-x}$O$_3$ films on Ge. The distances are in units of Angstroms. The superscripts refer to the layers in Figure 2.}
\label{table:2}
\end{center}
\end{table}

For all the films studied, polar rumpling of the oxygen-metal ions in the vertical direction are experimentally observed in agreement with results for strained STO on Si.\cite{kolpak2010interface,kumah2010atomic} For the BO$_2$ (B=Ti,Zr) layer closest to the interface, the amplitude of the rumpling decreases from 0.3 \AA{} for (SZTO x=1.0) to 0.25 \AA{}  for (SZTO x=0) with the oxygen ions moving down relative to the cation z-positions towards the Ge substrate. For the STO film, the oxygen displacements are negative (positive polarization) at the interface and switch sign in the surface SrO layer. In contrast, the (SZTO x=1.0) film has a uniform positive polarization in all the film layers.



 The DFT calculated structures of the SZTO/Ge interface are shown in Figure \ref{fig:bond-model} as a function of Zr-content. To compare the theoretical and experimental structures, the ionic buckling and layer spacings determined for the theoretically relaxed structures are shown in Table \ref{table:2}. The theoretical results are in good agreement with the experimentally measured structures.

\begin{figure}[ht]
\includegraphics[scale=0.45]{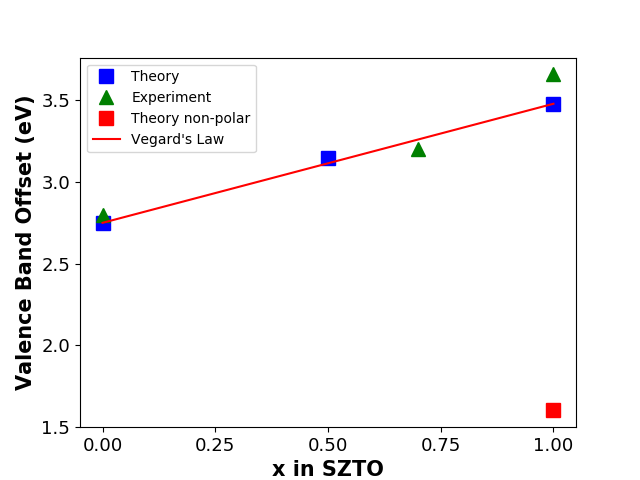} 
\caption{Comparison of DFT and experimental SrZr$_{x}$Ti$_{1-x}$O$_3$/Ge valence band offsets (VBO). The solid red line indicates the VBO calculated from Vegard's law. The DFT calculated VBO for an x=1.0 structure where rumpling within the oxide layers are suppressed (non-polar) is shown (red square) for comparison with the values for the VBO's determined for the fully converged structures (blue squares).The experimental band offsets for x=0, x=0.5 and x=1 are obtained from ref. \cite{chambers2017effects}, ref. \cite{jahangir2015band} and ref. \cite{lim2017structural} respectively. }

\label{fig:compare_bands_VBO}
\end{figure}

To elucidate the effects of the atomic-scale structure on the electronic properties of the films, the calculated valence band offsets of the theoretically fully relaxed structures are compared in Figure \ref{fig:compare_bands_VBO} as a function of Zr-content. The theoretical valence band offsets for the SZT interfaces are 2.75 eV, 3.15 eV and 3.48 eV for the x=0, x=0.5 and x=1.0 interfaces respectively. These values are in excellent agreement with XPS measurements of valence band offsets of 2.8 eV for STO/Ge\cite{chambers2017effects} and 3.66 eV for SZO/Ge\cite{lim2017structural} Due to the underestimation of the SZTO band gap by DFT,\cite{weston2015structural, Hybertsen1985First,yakovkin2004srtio}, calculated conduction band offsets cannot be directly compared with experiments. 


The evolution of the band offsets with Zr-content is consistent with the changes in the bulk band gap of SZTO as a function of Zr. However, to determine the role of the polar structural distortions on the valence band offsets, we compare the calculated band structure for a fully structurally relaxed (SZTO x=1.0)/Ge slab with a slab of similar composition where the anions are constrained to be in the same horizontal plane as the cations resulting in a non-polar structure. The non-polar constrained films are found to be less energetically stable by 1.75 eV. A comparison of the layer resolved electronic structure of the fully relaxed and constrained film is also shown in Figure \ref{fig:compare_bands_VBO}. The absence of the polar distortions in the SZTO films leads to a redistribution of charge in the bottom Zr layer to screen the interfacial charge resulting in a reduction in the valence band offset. The calculated valence band offset for the (SZTO x=1)/Ge structure which is constrained to have no polar distortions is found to be 1.60 eV compared to 3.48 eV for the fully converged (polar) (SZTO x=1) structure.

%
%


In conclusion, polar distortions are observed at the interface between SZTO films and Ge. The distort  ions are found to evolve with the distance from the interface and the Zr content. Using DFT, we find that suppressing the polar distortions leads to an electronic reconstruction which significantly modifies the interfacial band offsets. These results highlight the interplay between the atomic-scale structural properties of oxide-semiconductor interfaces and relevant technological properties such as band offsets.

This work was supported by the National Science Foundation (NSF) under award No. DMR1508530. This research used resources of the National Energy Research Scientific Computing Center, a DOE Office of Science User Facility supported by the Office of Science of the U.S. Department of Energy under Contract No. DE-AC02-05CH11231.Use of the Advanced Photon Source was supported by the U. S. Department of Energy, Office of Science, Office of Basic Energy Sciences, under Contract No. DE-AC02-06CH11357.

\bibliography{main}

\end{document}